\documentclass[reprint,
superscriptaddress,
 amsmath,amssymb,
 aps, physrev,
]{revtex4-2}

\usepackage{graphicx}
\usepackage{dcolumn}
\usepackage{bm}
\usepackage{hyperref}
\usepackage{color}

\begin{document}

\title{\textbf{Simultaneous calibration of rotation and phase errors in a single experiment}}

\author{Tien D. Nguyen}
 \email{dinh-tien.nguyen@polytechnique.edu}
 \affiliation{Faculty of Physics, Hanoi University of Science, VNU Hanoi 120401, Vietnam}
 \affiliation{Laboratoire de Physique de la Matière condensée, CNRS,
Ecole Polytechnique, Institut Polytechnique de Paris, 91120 Palaiseau, France}
\author{Hung Q. Nguyen}
 \email{hungngq@hus.edu.vn}
 \affiliation{Nano and Energy Center, Faculty of Physics, VNU University of Science, Vietnam National University, 120401, Hanoi, Vietnam}
 \affiliation{Institute for Quantum Technologies, VNU Technology and Innovation Park, Vietnam National University, 13151, Hanoi, Vietnam}

\date{\today}

\begin{abstract}
In weakly anharmonic qubits, coherent control errors take two generic forms---over/under-rotation and phase errors---whose suppression normally requires iterated experiments. We show that, for any symmetric $\pi/2$ pulse in the weak-driving regime, both follow from a single parametrization, $\mathcal{X}(\pi/2)=Z(\delta)X(\pi/2+\epsilon)Z(\delta)$, that ties the rotation error $\epsilon$ and the phase error $\delta$ directly to the system parameters. The parametrization enables DRAPE, a Ramsey-type protocol in which sweeping the phase-error correction reveals a crossing point that fixes both corrections at once. The phase correction is estimated with Heisenberg scaling while the rotation error saturates the standard quantum limit. We experimentally demonstrate DRAPE by calibrating a $\pi/2$ gate on the $|1\rangle\leftrightarrow |2\rangle$ transition of an IBM transmon, reducing the over-rotation from $0.997^\circ$ to $-0.007^\circ$ and the phase error from $2.52^\circ$ to $0.0052^\circ$ per gate, validated independently by phase- and rotation-error amplification protocols.
\end{abstract}

\maketitle

\textit{Introduction---}In weakly anharmonic systems such as superconducting qubits \cite{koch07}, control nonidealities, that is, a finite anharmonicity and a finite-duration drive, produce coherent (unitary) errors that, at the circuit level, take two generic forms: over/under-rotation errors and phase errors. Because two-level unitaries are universal \cite{RZBB94}, these errors appear in control schemes across Hilbert-space dimensions, including the qudit ladder gate set \cite{li25_recursivedrag,Wang25_HighEJEC,morvan21_rb}. While the physical error rate is most often identified with the intrinsic (nonunitary) decoherence of the qubit \cite{acharya_quantum_2025}, coherent gate errors must likewise be suppressed below a fault-tolerance threshold for quantum error correction to succeed \cite{barends14_logicgate,Preskill_1998}. Because solid-state processors drift between runs \cite{krantz19_guide,blais21}, this suppression must be recalibrated routinely, making the efficient and scalable characterization of coherent errors increasingly important as the qubit count rises. 

Quantum process and gate-set tomography describe coherent errors completely but at high cost \cite{cao24_gst}, while randomized benchmarking \cite{morvan21_rb} is resource-efficient yet returns an average gate fidelity that does not directly guide calibration. Calibration of single-qubit gates, most notably the $\pi/2$ gate, therefore relies on two amplified-error metrologies. Amplified phase error (APE) \cite{lucero10} first fixes a phase correction---through detuning, derivative removal of adiabatic gates (DRAG) \cite{motzoi09,gambetta11}, or virtual-$Z$ gates \cite{mckay17}---after which the drive amplitude is rescaled using amplified rotation error (ARE) \cite{sheldon2015}. Because rescaling the amplitude reintroduces a different phase error, the two steps must be iterated, converging only after several cycles \cite{hyyppa2024_reducing_leakage,chen16}. Both conceptually fall under the more general robust phase estimation (RPE) protocol \cite{kimmel15,rudinger17}, which extracts systematic amplitude and off-resonance errors at the Heisenberg limit, but with a dedicated circuit, ARE-like or APE-like, per parameter. 

In this Letter we show that both errors follow from a single, microscopically derived parametrization of the realized $\pi/2$ gate, in which the over/under-rotation $\epsilon$ and the phase error $\delta$ are tied explicitly to the drive and qubit parameters. This generalizes the empirical $ZXZ$ form of Ref.~\cite{lucero10}, which only describes phase error, by incorporating the rotation error. Furthermore, our analysis exposes the quantitative asymmetry of their interplay: in the weak-driving regime $\epsilon$ depends only on the drive amplitude, whereas $\delta$ depends on $\epsilon$, so correcting the phase leaves the rotation untouched while correcting the rotation shifts the phase. Building on this parametrization, we introduce DRAPE (DRAG-APE), a single Ramsey-type experiment that resolves and corrects $\epsilon$ and $\delta$ together: sweeping the DRAG coefficient exposes a crossing point at which both corrections are fixed simultaneously. We show that our protocol estimates the phase correction with Heisenberg scaling in the number of amplification cycles, while the rotation error saturates the standard quantum limit. 

We experimentally demonstrate DRAPE on a cloud-accessible IBM transmon \cite{alex20}, calibrating a $\pi/2$ gate of a qubit encoded in the $|1\rangle\leftrightarrow|2\rangle$ higher energy transition to underscore the protocol's applicability across the qudit ladder. From a single experiment, we extract both corrections and validate them independently against APE and ARE, suppressing the coherent errors by two to three orders of magnitude without iterative calibration.

\textit{Error parametrization---}We consider a qubit formed by two states $|j\rangle, |j+1\rangle$ of a weakly anharmonic oscillator of fundamental energy $\hbar\omega$ and anharmonicity $\hbar\Delta_2$, controlled by an in-phase and quadrature pulse $\Omega(t)=\Omega_x(t)-i\Omega_y(t)$; when $\Omega_y\propto\dot\Omega_x$, the pulse carries a derivative removal by adiabatic gate (DRAG) correction \cite{motzoi09, gambetta11}. The small parameter is $\Omega/\Delta_2$, i.e. a weakly driven qubit. In the drive frame, we perturbatively eliminate two nearest neighbor states $|j-1\rangle$ and $|j+2\rangle$ [see Supplemental Material (SM)]. This results in the effective Hamiltonian
\begin{align}\label{main_eqn:effective_qubit_Hamiltonian}
    H_{\text{eff}}(t) &={\lambda_j}\left[\frac{\Omega_x(t)}{2}\sigma^x_{j,j+1}+\frac{\Omega_y(t)}{2}\sigma^y_{j,j+1}\right]\notag\\
    &\quad+\left[\frac{\delta_q}{2}+\frac{\lambda_{j+1}^2-\lambda_{j-1}^2}{8\Delta_2}|\Omega(t)|^2\right]\sigma^z_{j,j+1}\notag\\
    &\quad+O(\Omega^3/\Delta_2^2).
\end{align}
Here, the $\sigma^z$ term includes two frequency shifts: the dynamical ac Stark shift, which scales with the drive strength $|\Omega(t)|^2$ and is suppressed by the nonlinearity $\Delta_2$, and a detuning $\delta_q$, whose microscopic origin---resonant two-level defects \cite{martinis18_TLSresonance}, quasiparticle tunneling \cite{catelani12_quasiparticletheory,riste13}, or background charge noise \cite{Christensen19_anomalouschargenoise}---is irrelevant here: these processes are slow compared with the gate time, so $\delta_q$ acts as a static detuning that requires calibration. Throughout, we denote $\zeta(t) = {\delta_q}+({\lambda_{j+1}^2-\lambda_{j-1}^2})|\Omega(t)|^2/{4\Delta_2}$ the qubit net frequency shift.

To see how coherent rotation and phase errors arise at the circuit level, it is instructive to first consider the simplest setting: a flat pulse with no DRAG correction, for which the effective Hamiltonian is time independent, $H_{\text{eff}}=\frac{1}{2}\left(\Omega\sigma^{x}_{j,j+1}+\zeta\sigma^{z}_{j,j+1}\right)$ (we set $\lambda_j=1$). The Hamiltonian describes a spin-$1/2$ particle in a fictitious field $\vec{R}=\left(\Omega,0,\zeta\right)$ about which the Bloch vector precesses at a rate $R=(\Omega^2+\zeta^2)^{1/2}$; weak driving implies $\zeta\ll\Omega$: the Stark contribution scales as $\Omega^2/\Delta_2$, and the drift $\delta_q$ is smaller still in practice. The net frequency shift $\zeta$ tilts $\vec{R}$ out of the equatorial ($xy$) plane by $\tan\alpha=\zeta/\Omega$, hence deflects the final Bloch vector by an azimuthal angle $\delta\approx\zeta/\Omega$ away from the $-\hat{y}$ axis; the angle $\delta$ is called a \textit{phase} error \cite{lucero10}. Indeed, to first order in $\alpha$, the tilted rotation takes precisely the form $Z(\delta)X(\pi/2)Z(\delta)$, with equal phases on both sides. The rotation angle, by contrast, is $RT_g\approx \Omega T_g(1+\frac{\zeta^2}{2\Omega^2})$: calibrating $\Omega T_g=\pi/2$, the detuning $\zeta$ perturbs the rotation only at second order, so a first-order \textit{rotation} error requires a miscalibrated amplitude $\Omega$. Conversely, rescaling $\Omega$ to correct the rotation shifts the phase error $\delta\approx\zeta/\Omega$. This static picture therefore heralds the one-directional interplay between the two errors.

To obtain the genuine error parametrization that can be employed for error calibration, one must consider the full time-dependent problem. To this end, we move into the toggling frame defined by $V_{\text{Tog}}(t) = \exp[-i\frac{\theta(t)}{2}\sigma^x_{j,j+1}]$, where $\theta(t)=\lambda_j\int_0^t \Omega_x(\tau)d\tau$ is the rotated angle. The transformed Hamiltonian, $H_{\text{Tog}}=V^{\dagger}_{\text{Tog}}H_{\text{eff}}V_{\text{Tog}} - iV^\dagger_{\text{Tog}}\dot{V}_{\text{Tog}}$, reads
\begin{align}\label{eqn:toggling_frame_Hamiltonian}
    H_{\text{Tog}}(t) &= \frac{1}{2}\left[\lambda_j{\Omega_y(t)}\cos\theta+\zeta(t)\sin\theta\right]\sigma^y_{j,j+1}\notag\\
    &\quad+\frac{1}{2}\left[-\lambda_j{\Omega_y(t)}\sin\theta+\zeta(t)\cos\theta\right]\sigma^z_{j,j+1}
\end{align}
The $H_{\text{Tog}}$ Hamiltonian describes residual errors in the toggling frame \cite{jesus2025_analyticalblueprintxgates, togglingframe}. The corresponding residual unitary $\tilde{U}(T)$ is (see SM for derivation)
\begin{align}
    \tilde{U}(T)&\approx\exp\left[-i\frac{\delta_{y}(T)}{2}\sigma^y_{j,j+1}\right]\exp\left[-i\frac{\delta_{z}(T)}{2}\sigma^z_{j,j+1}\right],\notag\\
    &=Y(\delta_y)Z(\delta_z).
\end{align}
In the drive frame, where the qubit is encoded, the total unitary of the system is the product of the ideal operation and the residual unitary, $U(T) = V_{\text{Tog}}(T)\tilde{U}(T)$. We call $\delta_y$ and $\delta_z$ the residual errors.

We now consider two conditions: a symmetric pulse of duration $T_g$, $\Omega_x(T_g-t)=\Omega_x(t)$ with antisymmetric quadrature $\Omega_y(T_g-t)=-\Omega_y(t)$, but miscalibrated as $\Omega_x(t)=\Omega_0(1+\eta)f(t)$ with a small fractional amplitude $\eta\ll 1$, which represents the primary source of under/over-rotation errors, and an intended rotation angle of $\pm\pi/2$, implying $\theta(T_g-t)\approx\pm\frac{\pi}{2}(1+\eta)-\theta(t)$. Under these conditions, the toggling frame Hamiltonian admits an approximate time-reversal symmetry exchanging $\sigma^y\leftrightarrow\sigma^z$ under $t\to T_g-t$, so that after $T_g$ the residual errors become approximately equal, $\delta_y(T_g)\approx\delta_z(T_g)\equiv\delta$. Most importantly, the $\sigma^y$-type error propagates to the end of the interaction as a $\sigma^z$ \textit{phase} error, $XY(\delta)\approx Z(\delta)X$. We arrive at the parametrization of an erroneous $\mathcal{X}(\pi/2)$
\begin{align}\label{eqn:parametrization}
    \mathcal{X}(\pm\pi/2) &=V_\text{Tog}(T_g)\tilde{U}(T_g),\\
    &= Z(\delta)X(\pm(\pi/2+\epsilon))Z(\delta),
\end{align}
where $\epsilon$ is the over/under-rotation error and $\delta$ is the phase error; both unambiguously depend on the drive and qubit parameters as,
\begin{align}\label{eqn:coherent_error_parameters_a}
    \epsilon &= \eta\lambda_j\Omega_0\int_{0}^{T_g} f(t) dt,\\
    \delta &= \int_0^{T_g}\left[-\lambda_j{\Omega_y(t;\epsilon)}\sin\theta(t;\epsilon)+\zeta(t;\epsilon)\cos\theta(t;\epsilon)\right]dt.\label{eqn:coherent_error_parameters_b}
\end{align}
The first condition encompasses common pulse envelopes (truncated Gaussian, raised-cosine/Hann, or symmetry-constrained optimal control pulses); the second is natural since a $\pi/2$ gate and an arbitrary-angle $Z$ gate together form a qubit \cite{mckay17} and qudit \cite{li25_recursivedrag} native gate set, and if the $Z$ gate is virtual (thus assumed error-free) only the $\pi/2$ gate needs optimization. 

In the weak driving regime, the inter-dependence between $\epsilon$ and $\delta$ is one-directional, as anticipated by the static picture: $\epsilon$ depends only on the fractional amplitude, whereas $\delta$ depends on $\epsilon$ via the ac Stark shift strength and the rotated angle. Correcting $\delta$ thus does not shift $\epsilon$, while correcting $\epsilon$ shifts $\delta$; the opposite backaction enters strictly from second order, where the off-diagonal elements of the effective Hamiltonian are dressed by virtual two-photon processes. In practice, this interplay explains why existing iterative calibration loops \cite{hyyppa2024_reducing_leakage,chen16} for weakly driven qubits converge only after a few cycles. In the recently proposed \cite{jesus2025_analyticalblueprintxgates} and experimentally realized \cite{gao2025ultrafastsinglequbitgates} strong-driving regime, $\Omega/\Delta_2\approx 0.33$, this backaction is no longer negligible; a compact parametrization there is beyond the scope of this work.

\begin{figure*}
\centering
\begin{minipage}{3.404in}
  \includegraphics{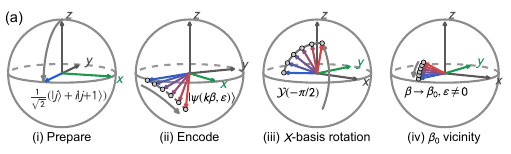}\\[-0.5ex]
  \includegraphics{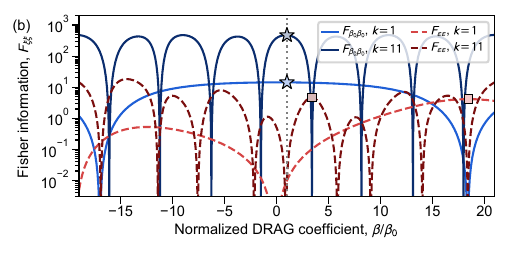}
\end{minipage}
\begin{minipage}{3.404in}
  \includegraphics{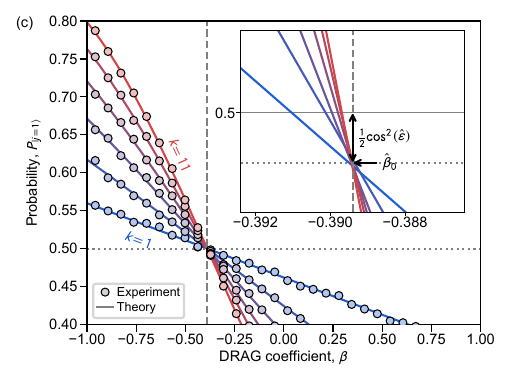}
\end{minipage}
\caption{DRAPE (DRAG-APE) protocol. (a) Bloch-sphere geometry of the gate sequence: an imperfect $|-y\rangle$ state is prepared, $k$ pseudo-identity pairs encode $\epsilon$ and $\delta$ on its polar and azimuthal angles, and a final $y$-rotation converts the accumulated phase into a measurable population. (b) Per-shot Fisher information versus the DRAG sweep $\beta/\beta_0$: $F_{\beta_0\beta_0}$ (solid blue) and $F_{\epsilon\epsilon}$ (dashed red) for $k=1$ and $11$. At the crossing point $\beta_0$ (dotted line) $F_{\beta_0\beta_0}$ is maximal while $F_{\epsilon\epsilon}$ is minimal; squares mark the $F_{\epsilon\epsilon}$ peak. (c) DRAPE measurement on the $|j=1\rangle\leftrightarrow|j+1=2\rangle$ transition of an IBM transmon: $P_{|j=1\rangle}$ versus $\beta$ for $k=1,\dots,11$; solid curves are fits to Eq.~(\ref{eqn:probability_DRAPE}). Inset: zoom on the crossing locus that simultaneously determines $\hat\epsilon$ and $\hat\beta_0$.}
\label{fig:drape}
\end{figure*}

\textit{DRAPE protocol---}Enabled by the parametrization of coherent errors, we propose a protocol that detects and corrects $X(\pi/2)$ over/under-rotation $\epsilon$ and phase error $\delta$ simultaneously. No calibration loop is required: the protocol reveals a crossing point at which both corrections are determined.

Consider the action of a composite unitary $U$ on the ground state $|j\rangle$, given by
\begin{align}
    U = \mathcal{Y}(\pi/2)\left[\mathcal{X}(-\pi/2)\mathcal{X}(\pi/2)\right]^k\mathcal{X}(\pi/2),
\end{align}
where $\mathcal{Y}=Z(\pi/2)\mathcal{X}Z(-\pi/2)$. Fig.~\ref{fig:drape}(a) depicts the geometry of $U$ on the Bloch sphere. The first $\mathcal{X}(\pi/2)$ prepares a near-equal superposition $|-y\rangle=\frac{1}{\sqrt{2}}(|j\rangle-i|j+1\rangle)$; $k$ pseudo-identity operations then encode $\epsilon$ and $\delta$ on its polar and azimuthal angles; a final orthogonal rotation $\mathcal{Y}$ maps the accumulated $k\delta$ phase to a measurable population. The composite $U$ resembles the Ramsey sequence of the APE protocol \cite{lucero10}, with the final rotation axis parked at $\phi=\pi/2$ (largest probability slope) to maximize the interference signal. One can estimate $\epsilon$ and $\delta$ by measuring $P_{|j\rangle}=|\langle j|U(\epsilon,\delta)|j\rangle|^2$,
\begin{align}\label{eqn:probability_DRAPE}
    P_{|j\rangle}=1-\frac{\cos^2\epsilon}{2}[(u+v+f\cos\delta)^2+(v+f\sin\delta)^2],
\end{align}
where the coefficients $u,v,f$, given in the SM, are combinations of the Chebyshev polynomial of the second kind $S_k(\cos\phi)=\sin[(k+1)\phi]/\sin\phi$ evaluated at $\cos\phi=\cos^2\delta+\sin\epsilon\sin^2\delta$. Experimentally, phase error $\delta$ is probed through a control knob, for example pulse detuning \cite{chen16}, arguments of virtual $Z$ gates \cite{mckay17}, or DRAG quadrature component \cite{lucero10}. In this Letter, we consider the DRAG correction $\beta$, on which $\delta$ depends linearly [cf. Eq.~(\ref{eqn:coherent_error_parameters_b})],
\begin{align}\label{eqn:the_DRAPE_slope}
    \delta(\beta) &= (\beta_0-\beta)\frac{\lambda_j}{\Delta_2}\int_0^{T_g}\dot\Omega_x(t)\cos\theta(t)dt,
\end{align}
where $\beta_0$ is the zero of $\delta(\beta)$, i.e. the optimal DRAG correction for phase error. We term our protocol DR(ag)-APE, or DRAPE; we chose DRAG because the linearity between $\delta$ and $\beta$ simplifies the analysis, and similar protocols follow analogously for the other phase-correction knobs.

To illustrate how our protocol works, it is instructive to expand $P_{|j\rangle}$ to first order in $\beta$ at arbitrary $\epsilon$. Denoting $\chi\equiv\frac{\lambda_j}{\Delta_2}\int_0^{T_g}\dot\Omega_x(t)\cos\theta(t)dt$, we have
\begin{align}\label{eqn:probability_linear}
    P^{\text{lin}}_{|j\rangle} = 1-\cos^2\epsilon\left[\frac{1}{2}+\chi\beta_0(1-\frac{\beta}{\beta_0})(1+k(1-2\sin\epsilon))\right].
\end{align}
As $\beta\to\beta_0$, the phase error $\delta\to0$, and Eq.~(\ref{eqn:probability_linear}) reveals a crossing point $\beta_0$ where the $k$-dependence of the probability vanishes, allowing an estimate $\hat\beta_0$ of the phase-error correction. While the sign of $\epsilon$ is ambiguous exactly at $\beta_0$, the proper estimate $\hat\epsilon$ can be recovered by moving away from the crossing point and analyzing the $k$-dependent slopes, which encode both its magnitude and sign.

In going from a single-parameter to a joint two-parameter estimation problem, it is natural to ask what trade-off has to be made. We answer this with the Fisher information extracted from $N$ shots of a DRAPE circuit, $F_{\xi\xi}=N(\partial_\xi P_{|j\rangle})^2/[P_{|j\rangle}(1-P_{|j\rangle})]$, which bounds the estimates through $\sigma(\hat\xi)\ge 1/\sqrt{F_{\xi\xi}}$, $\xi\in\{\beta_0,\epsilon\}$ (SM). As $\beta\to\beta_0$ and $\epsilon\to 0$, $F_{\delta\delta}$ approaches $4N(k+1)^2$ while $F_{\epsilon\epsilon}$ approaches $4N\epsilon^2$; since $\delta$ and $\beta_0$ differ by the fixed scale $\chi$, $F_{\beta_0\beta_0}=\chi^2 F_{\delta\delta}$ carries the same $(k+1)^2$ dependence, and because $k$ and $N$ are independent, DRAPE achieves Heisenberg scaling for estimating $\beta_0$. Meanwhile, $F_{\epsilon\epsilon}$ is bounded above at fixed $N$: $\hat\epsilon$ can be estimated simultaneously, but only at the standard quantum limit (SQL), and increasing $k$ merely moves the $F_{\epsilon\epsilon}$ peak toward $\beta_0$ at fixed height. We verify these claims numerically in Fig.~\ref{fig:drape}(b) at $\epsilon=0.05$ and $\chi\beta_0=-0.02$: $F_{\beta_0\beta_0}$ is maximal at the crossing, improving by $\approx 33$ from $k=1$ to $11$ (asymptotically $(k{+}1)^2/4=36$), while $F_{\epsilon\epsilon}$ dips to a minimum.

Importantly, the Heisenberg scaling of $\hat\beta_0$, inherited from the constructive interference of a Ramsey sequence, is not compromised by learning $\hat\epsilon$ simultaneously; the gain is achieved by simply using the correct error parametrization. In contrast, RPE \cite{kimmel15,rudinger17} attains Heisenberg scaling for \textit{both} errors, but with a separate circuit for each in a sequential manner; whether a \textit{single} circuit can attain Heisenberg scaling for both errors at once remains, to the best of our knowledge, an open question.

\begin{figure}
\includegraphics{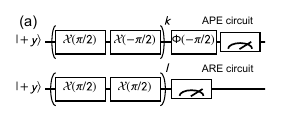}
\includegraphics{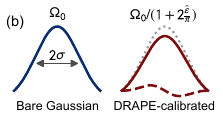}\\[-1.0ex]
\includegraphics{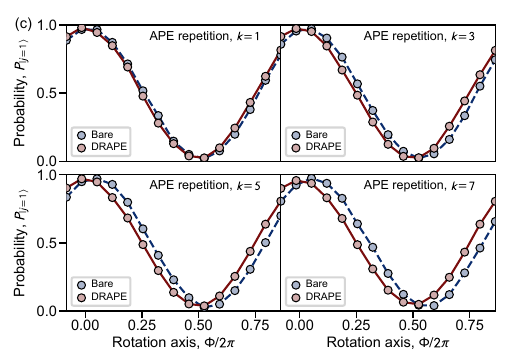}\\[-0.5ex]
\includegraphics{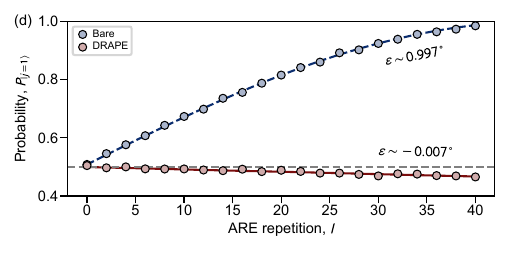}
\caption{\label{fig:aae_ape}Validation of the DRAPE corrections. (a) The APE and ARE circuits. (b) Pulse shapes: bare Gaussian versus DRAPE-calibrated Gaussian, whose amplitude is rescaled to cancel $\hat\epsilon$ and to which a DRAG quadrature $\Omega_y=-(\hat\beta_0/\Delta_2)\dot\Omega_x$ is added to cancel the phase error. (c) APE result: $P_{|j=1\rangle}$ versus the final-axis azimuth $\phi$ for $k=1,3,5,7$; the Ramsey fringes fan out with $k$ before correction and collapse onto a single curve after. (d) ARE result: $P_{|j=1\rangle}$ after $\ell$ repetitions of $\mathcal{X}(\pi/2)\mathcal{X}(\pi/2)$; the population drifts from the ideal $1/2$ before correction and stays near $1/2$ after.}
\end{figure}

\textit{Experimental demonstration---}We implement DRAPE on transmon qubit 109 of the \textit{brisbane} processor, made cloud-accessible by the IBM Quantum Experience program \cite{alex20}. Our data were obtained between 2024 and early 2025, before the deprecation of the Qiskit Pulse module. The qubit is encoded in the $|j=1\rangle\leftrightarrow|j+1=2\rangle$ transition, separated from the leakage subspace by an anharmonicity $|\Delta_2|/2\pi=307$ MHz. Our uncalibrated control is a truncated $T_g=20$ ns Gaussian pulse $\Omega_x(t)=\Omega_0f_G(t)$ [see End Matter for the detailed waveform]. The drive amplitude $\Omega_0/2\pi=40$ MHz is determined by a Rabi oscillation experiment, and each circuit is executed for $N=10^4$ shots. We simulate the system using the four-level RWA Hamiltonian of a driven nonlinear oscillator \cite{gambetta11}, with one free parameter, $\delta_q$, tracking the qubit frequency fluctuations.

In Fig.~\ref{fig:drape}(c), we show the observed $P_{|j=1\rangle}$ population versus $\beta$ during a DRAPE experiment for $k$ from $1$ to $11$ (scattered points). Our simulation and the fit to Eq.~(\ref{eqn:probability_DRAPE}) agree numerically, validating an analysis accurate to first order in $\Omega/\Delta_2$; the only discrepancy, between the theoretical $\chi^{\text{th}}=4.99\times 10^{-2}$ and the empirical $\chi^\text{em}=5.03\times 10^{-2}$, is consistent with a second-order correction. Zooming in the crossing point region, the probability of measuring $|j=1\rangle$ exhibits a clear linear relation with $\beta$, as predicted by Eq.~(\ref{eqn:probability_linear}). Using DRAPE theory, we extract $\hat\epsilon = (0.997 \pm 0.395)^\circ$ and $\hat\beta_0=-0.3894 \pm 0.0002$. Our estimator probes only the vicinity of $\hat\beta_0$: away from it, $\sigma(\hat\epsilon)$ decreases with the larger Fisher information, but the estimate becomes biased by higher-order corrections in $\Omega/\Delta_2$.

The over-rotation is expected: our $\pi/2$ amplitude is obtained by linearly downscaling the Rabi-calibrated $\pi$-pulse amplitude, a procedure known to yield systematic over-rotations \cite{lazar23}. DRAPE theory predicts the optimal correction by solving $\delta(\beta)=0$, 
\begin{align}
    \beta_0=\frac{\lambda_{j+1}^2-\lambda_{j-1}^2}{4\lambda_j^2}+\frac{\Delta_2\delta_q}{\lambda_j^2}\frac{\int_0^{T_g}\sin\theta dt}{\int_0^{T_g}\Omega_x^2\sin\theta dt}.
% NOTE: sign of both terms corrected to match the j=0 limit (beta_0=+1/2) and the SM sign conventions.
% Please verify the extracted |delta_q|/2pi = 0.466 MHz was obtained with THIS sign of the formula.
\end{align}
The first term corrects the ac Stark shift and depends only on the qubit's energy structure; the second, pulse-dependent term corrects small qubit-frequency drifts, $\delta_q\lesssim \Omega^2/\Delta_2$ (End Matter). For our pulse shape this yields a drift $|\delta_q|/2\pi\approx 0.466$ MHz, well within the $\approx 2$ MHz charge dispersion of the $|j=1\rangle\leftrightarrow|j+1=2\rangle$ transition at $E_J/E_C=37$.

We validate $\hat\epsilon$ and $\hat\beta_0$ through their respective optimal protocols, amplified rotation error (ARE) \cite{sheldon2015}, or $N$-pulse amplification \cite{lazar23}, and amplified phase error (APE) \cite{lucero10}, whose circuits are shown in Fig.~\ref{fig:aae_ape}(a). Both start from $|-y\rangle$. ARE repeats the pair $\mathcal{X}(\pi/2)\mathcal{X}(\pi/2)$, ideally mapping $|-y\rangle\leftrightarrow|+y\rangle$, $\ell$ times with a fixed rotation axis; the deviation from $P_{|j\rangle}=0.5$ reveals the coherently amplified rotation error $\epsilon$. APE, similar to the DRAPE sequence, mostly amplifies the phase error $\delta$, probed by varying the final rotation axis. In each experiment, we compare two cases: bare Gaussian pulse and DRAPE-calibrated pulse, where we apply both amplitude and phase correction, 
\begin{align}
    \Omega_x(t) &= \frac{\Omega_0}{1+2\hat\epsilon/\pi}f_G(t),\\
    \Omega_y(t) &= -\frac{\hat\beta_0}{\Delta_2}\dot\Omega_x(t).
\end{align}
In the APE experiment [Fig.~\ref{fig:aae_ape}(c)], the Ramsey fringes shift linearly with $k$ before correction and align after, indicating phase errors suppressed from $\delta=2.521^\circ$ to $0.0052^\circ$. In the ARE experiment [Fig.~\ref{fig:aae_ape}(d)], rotation errors accumulate coherently with the number of repetitions $\ell$ before correction, while after correction the population stabilizes near $0.5$, reducing a $0.997^\circ$ over-rotation to a $-0.007^\circ$ under-rotation. Both measurements independently validate the DRAPE estimates without an iterative calibration loop.

\textit{Conclusion---}In the weak driving regime of a weakly anharmonic qubit, we have shown that the two generic coherent errors of a $\pi/2$ rotation, over/under-rotation $\epsilon$ and phase error $\delta$, follow from a single circuit-level parametrization, $\mathcal{X}(\pi/2)=Z(\delta)X(\pi/2+\epsilon)Z(\delta)$, which reveals the asymmetry at play: rescaling the drive amplitude to correct $\epsilon$ shifts the optimal solution for $\delta$, ordinarily demanding a feedback calibration loop. DRAPE avoids this loop by exposing a crossing point that fixes both corrections at once, achieving Heisenberg scaling for the phase estimate and the standard quantum limit for the rotation estimate in a single experimental run. We demonstrated the protocol on the $|1\rangle\leftrightarrow|2\rangle$ transition of a superconducting transmon and validated the extracted corrections independently against ARE and APE. Two questions stand out: whether a single circuit can reach the Heisenberg limit for both errors, and how the parametrization extends into the strongly driven regime, where the backaction of $\delta$ on $\epsilon$ is no longer negligible.

\begin{acknowledgments}
We would like to thank Nguyen V. K. Huynh, Linh N. T. Trinh, Duy V. Nguyen, and Le Bin Ho for helpful discussions. We thank IBM Quantum for providing generous access to their quantum hardware infrastructure and their members of technical staff for their valuable support. NQH would like to acknowledge the funding from Vietnam National University Hanoi under Grant Number QG.24.105.
\end{acknowledgments}

\section*{End Matter}
\appendix
\section{DRAG correction for phase errors}\label{appendix:DRAGcorrection}
We introduce the DRAG correction $\Omega_y=-\beta{\dot\Omega_x}/{\Delta_2}$ and write the ac-Stark coefficient as $\mu=(\lambda_{j+1}^2-\lambda_{j-1}^2)/4\Delta_2$. To leading order, the phase error $\delta$ is
\begin{align}  
    \delta &= \int_0^{T_g}\left[-\lambda_j\beta\frac{\dot\Omega_x}{\Delta_2}\cos\theta+\zeta(t)\sin\theta\right]dt,\\
    &= -\frac{\beta\lambda_j}{\Delta_2}\int_0^{T_g}\dot\Omega_x\cos\theta dt + \int_0^{T_g} (\delta_q+\mu\Omega_x^2)\sin\theta dt.
\end{align}
The $\sigma^y\leftrightarrow\sigma^z$ time-reversal symmetry holds even when we introduce the quadrature as $\Omega_y$ is anti-symmetric under $t\to T_g-t$. For any pulse that vanishes at the boundaries $\Omega_x(0)=\Omega_x(T_g)=0$, the suppressed phase error is achieved by setting $\delta=0$. This gives the optimal DRAG coefficient for phase error,
\begin{align}
    \beta_0 = \frac{\lambda^2_{j+1}-\lambda^2_{j-1}}{4\lambda_j^2}+\frac{\delta_q\Delta_2}{\lambda_j^2}\frac{\int_0^{T_g}\sin\theta(t)dt}{\int_0^{T_g}\Omega^2_x(t)\sin\theta(t)dt}.
\end{align}
We identify the first term with the DRAG correction for a pure dynamical ac Stark shift, $(\lambda^2_{j+1}-\lambda^2_{j-1})/4\lambda_j^2$, which is independent of pulse parameters (shape, duration, or amplitude) and depends only on the structure of the qubit transition. Substituting $j=0$ and $\lambda_{-1}=0$, we recover the optimal DRAG phase error correction $\beta_0=1/2$ for a qubit encoded in the $|0\rangle\leftrightarrow|1\rangle$ transition of a transmon circuit \cite{martinis14_sigmaz,lucero10,chen16}. The second term, on the other hand, is pulse-dependent and $\delta_q$-dependent. Because of the uniqueness of $\beta_0$, we can always ensure a DRAG correction for an off-resonantly driven qubit, as long as $\delta_q\lesssim \Omega^2/\Delta_2$. Beyond this limit, a DRAG correction can still be found, but the weak drive assumption may be violated as $\delta_q\Delta_2$ becomes large (i.e. a large quadrature component). A pulse detuning or a $Z$ correction at the same order will be more appropriate to tackle phase errors. The truncated Gaussian envelope used in our experiments reads
\begin{align}\label{eqn:Gaussian_envelope}
    f_G(t) = \begin{cases}
        \frac{e^{-(t-T_g/2)^2/2\sigma^2}-e^{-T_g^2/8\sigma^2}}{1-e^{-T_g^2/8\sigma^2}},& 0\leq t\leq T_g\\
        0,& \text{otherwise}
    \end{cases}
\end{align}

\bibliography{references}

\clearpage
\begin{center}
    \textbf{\large Supplemental Material for ``Simultaneous calibration of rotation and phase errors in a single experiment''}
\end{center}
\setcounter{equation}{0}
\renewcommand{\theequation}{S\arabic{equation}}

\section*{Time-dependent Schrieffer-Wolff transformation}\label{appendix:SWperturbation}
We consider a weakly anharmonic oscillator of fundamental energy $\hbar\omega$ and anharmonicity $\hbar\Delta_2$, driven by an external field $\Omega(t)=\Omega_x(t)-i\Omega_y(t)$. In the frame rotating at $\omega_d=\omega-\delta_d$ (ignoring fast-oscillating terms at $\pm\omega_d$), the system Hamiltonian reads \cite{gambetta11}
\begin{align}
    H(t) = &\sum_{j}(j\delta_d+\Delta_j)|j\rangle\langle j|\notag\\
    &+\frac{1}{2}\sum_{j=1}^{d-1}\lambda_{j-1}[\Omega_x(t)\sigma_{j-1,j}^x+\Omega_y(t)\sigma^y_{j-1,j}],
\end{align}
where $\Delta_j=j\Delta_2(j-1)/2$ is the anharmonicity for the $j$-th level, $\delta_d=\omega-\omega_d$ is the drive detuning, and $\lambda_j$ weigh the relative strength between $|j\rangle\leftrightarrow|j+1\rangle$ transitions. In circuit QED, since the qubit is capacitively coupled to a readout resonator, $\lambda_j\neq\sqrt{j+1}$ \cite{gambetta11}; we take them as fitting parameters. We fit them directly from the Rabi oscillation of each transition $j\leftrightarrow j+1$. 

We wish to derive an effective Hamiltonian that captures the dynamics of the two driven states $|j\rangle, |j+1\rangle$ when their transition is on resonance, $\delta_d=-j\Delta_2$. To this end, we truncate the Hamiltonian to a four-level one, including the adjacent states $|j-1\rangle, |j+2\rangle$. The truncated Hamiltonian reads
\begin{align}
    H_4(t)&=H_0+V_\text{bd}(t)+V_\text{bod}(t)\notag\\
    &=\frac{1}{2}\delta_q\sigma^z_{j,j+1}+\Delta_2{\Pi^\ell}\notag\\
    &\quad+\lambda_j\left[\frac{\Omega_x(t)}{2}\sigma^x_{j,j+1}+\frac{\Omega_y(t)}{2}\sigma^y_{j,j+1}\right]\notag\\
    &\quad+\sum_{k=j\pm 1}\lambda_k\left[\frac{\Omega_x(t)}{2}\sigma^x_{k,k+1}+\frac{\Omega_y(t)}{2}\sigma^y_{k,k+1}\right]
\end{align}
where $\Delta_2<0$ following transmon convention and $\delta_q$ is the qubit frequency drift, included phenomenologically. We have followed the Schrieffer-Wolff (SW) transformation in \cite{blais21}, where the Hamiltonian is partitioned into $H_0, V_\text{bd}$ that do not couple states in different subspaces, while $V_\text{bod}$ does. The different subspaces, qubit and leakage, are defined by the projectors 
\begin{align*}
    \Pi^{\text{q}} &= |j\rangle\langle j|+|j+1\rangle\langle j+1|,\\
    \Pi^\ell &= |j-1\rangle\langle j-1|+|j+2\rangle\langle j+2|.
\end{align*}
We now find a unitary transformation $U(t)=\exp{S(t)}$ that decouples the qubit subspace from the leakage subspace. This is obtained by truncating the BCH expansion of the transformed Hamiltonian $H'(t)$
\begin{align}\label{eqn:BCH_expansion_SW}
    H'(t)=U(t)H_4(t)U^\dagger(t)+i\dot{U}(t)U^\dagger(t)
\end{align}
at desired order. To keep track of the expansion, we formally expand $H_4$ and $S$ in power series
\begin{equation}\label{eqn:formal_expansion_SW}
\begin{aligned}
    H_4 &= H_4^{(0)}+\kappa H_4^{(1)} + \kappa^2H_4^{(2)}+\dots \\
    S   &= \kappa S^{(1)}+\kappa^2 S^{(2)}+\dots
\end{aligned}
\end{equation}
where $\kappa\sim\Omega/\Delta_2$ is the small parameter. The Schrieffer-Wolff transformation is found by inserting Eq.~(\ref{eqn:formal_expansion_SW}) into Eq.~(\ref{eqn:BCH_expansion_SW}) and collecting terms at each order $\kappa^n$
\begin{align}
H' &= H_4^{(0)} + \kappa\left[H_4^{(1)}+[S^{(1)}, H_4^{(0)}]+i\dot{S}^{(1)}\right] \notag\\
&+\kappa^2\left[ [S^{(1)}, H_4^{(1)}]+[S^{(2)}, H_4^{(0)}] \right. \notag\\
&\left. +\frac{1}{2}[S^{(1)},[S^{(1)},H_4^{(0)}]]+\frac{1}{2}[S^{(1)},i\dot{S}^{(1)}] \right]+O(\kappa^3)
\end{align}
We iteratively solve for $S^{(k)}$ and $H^{'(k)}$ by requiring they are block off-diagonal and block diagonal, respectively, at all orders.

\noindent\textbf{Zeroth order.} $H^{'(0)}=\frac{1}{2}\delta_q\sigma^z_{j,j+1}+\Delta_2\Pi^\ell$. The qubit subspace is near-degenerate at $\delta_q$ energy, while the leakage subspace is at $\Delta_2$.

\noindent\textbf{First order.} To first order in $\kappa$, $S^{(1)}$ is determined by
\begin{align}
    [S^{(1)}, H_0]+i\dot{S}^{(1)}=-V_\text{bod},
\end{align}
which, to leading order in $1/\Delta_2$,
\begin{align}\label{eqn:S1_first_order}
S^{(1)} = &\sum_{k\in\{j\pm 1\}}\frac{i\lambda_k}{2\Delta_2}\left[\nu_k\Omega_x\sigma^y_{k,k+1}-\nu_k\Omega_y\sigma^x_{k,k+1}\right],
\end{align}
where $\nu_{j-1}=+1$ and $\nu_{j+1}=-1$. The effective Hamiltonian at the first order is the Rabi drive term
\begin{align}
    H^{'(1)} = &\lambda_j\left[\frac{\Omega_x}{2}\sigma^x_{j,j+1}+\frac{\Omega_y}{2}\sigma^y_{j,j+1}\right].
\end{align}
\noindent\textbf{Second order.} We choose $S^{(2)}$ to cancel the block-off-diagonal part $[S^{(1)}, V_\text{bd}]+[S^{(2)}, H_0]=0$. The remaining term reads 
\begin{align}
    H^{'(2)} &= [S^{(1)}, V_\text{bod}] + \frac{1}{2}[S^{(1)}, [S^{(1)}, H_0]] + \frac{1}{2}[S^{(1)}, i\dot{S}^{(1)}],\notag\\
    &=\frac{1}{2}[S^{(1)}, V_\text{bod}],\notag\\
    &=\sum_{k\in\{j\pm 1\}}\frac{\nu_k\lambda_k^2}{4\Delta_2}|\Omega|^2\sigma^z_{k,k+1},
\end{align}
where we have used Eq.~(\ref{eqn:S1_first_order}) and the identity $[a\sigma^x_{m,n}+b\sigma^y_{m,n}, c\sigma^x_{m,n}+d\sigma^y_{m,n}]=2i(ad-bc)\sigma^z_{m,n}$. Projecting $H^{'(2)}$ into $\Pi^q$ and combining with $H^{'(1)}$, we obtain the effective time-dependent qubit Hamiltonian,
\begin{align}\label{eqn:effective_qubit_Hamiltonian}
    H_{\text{S-W}}^{j,j+1}(t) =\lambda_j&\left[\frac{\Omega_x(t)}{2}\sigma^x_{j,j+1}+\frac{\Omega_y(t)}{2}\sigma^y_{j,j+1}\right]\notag\\
    &+\left[\frac{\delta_q}{2}+\frac{\lambda_{j+1}^2-\lambda_{j-1}^2}{8\Delta_2}|\Omega(t)|^2\right]\sigma^z_{j,j+1}.
\end{align} 
We identify two sources of phase errors: the static detuning $\delta_q$ and the dynamical ac Stark shift, which is proportional to $\Omega^2/\Delta_2$ due to the presence of leakage states. We note that the perturbation series is in powers of $\kappa\sim\Omega/\Delta_2$ while the unperturbed Hamiltonian has energy scale of $\Delta_2,\delta_q$; thus the effective Hamiltonian goes in powers of $\kappa$ times $\Delta_2$. 

\section*{Toggling-frame Hamiltonian and Magnus expansion}\label{appendix:toggling_frame_Magnus_expansion}
To understand how the second part of Hamiltonian (\ref{eqn:effective_qubit_Hamiltonian}) creates coherent phase errors, we move to the toggling frame defined by the ideal operation \cite{jesus2025_analyticalblueprintxgates}
\begin{align}
    V_{\text{Tog}}(t) = \exp\left[-i\frac{\theta(t)}{2}\sigma^x_{j,j+1}\right],    
\end{align}
where $\theta(t)=\lambda_j\int_0^t \Omega_x(\tau)d\tau$ is the ideal rotated angle at time $t$. We denote the net energy shift as $\zeta(t)$,
\begin{align}
    \zeta(t) = {\delta_q}+\frac{\lambda_{j+1}^2-\lambda_{j-1}^2}{4\Delta_2}|\Omega(t)|^2.
\end{align}
The toggling frame Hamiltonian is 
\begin{align}
    H_{\text{Tog}} &= V^{\dagger}_{\text{Tog}}H_{\text{S-W}}^{j,j+1}V_{\text{Tog}} - iV^\dagger_{\text{Tog}}\dot{V}_{\text{Tog}}\notag\\
    &=\frac{1}{2}\left[\lambda_j{\Omega_y(t)}\cos\theta+\zeta(t)\sin\theta\right]\sigma^y_{j,j+1}\notag\\
    &+\frac{1}{2}\left[-\lambda_j{\Omega_y(t)}\sin\theta+\zeta(t)\cos\theta\right]\sigma^z_{j,j+1},
\end{align}
which, by construction, describes the coherent error dynamics in the toggling frame. In the toggling frame, the state $|\tilde{\psi}(t)\rangle$ evolves as 
\begin{align}
    i\partial_t|\tilde{\psi}(t)\rangle = H_{\text{Tog}}(t)|\tilde{\psi}(t)\rangle,
\end{align}
whose formal solution is the time-ordered exponential
\begin{align}
\tilde{U}(T)=\mathcal{T}\exp\left[-i\int_0^T H_{\text{Tog}}(\tau)d\tau\right].
\end{align} 
In the drive frame, the unitary evolution is $U(T)=V_{\text{Tog}}(T)\tilde{U}(T)$, i.e. a multiplicative error. In fact, we can use the Magnus expansion to express $\tilde{U}(T)$ as an exponential of a series, $\tilde{U}(T)=\exp\left[-i\sum_{k=1}^\infty \mathcal{Z}_k(T)\right]$, where the first two terms are
\begin{align}
    \mathcal{Z}_1(T) &= \int_0^T H_{\text{Tog}}(\tau)d\tau,\\
    \mathcal{Z}_2(T) &= -\frac{i}{2}\int_0^T d\tau_1\int_0^{\tau_1}d\tau_2[H_{\text{Tog}}(\tau_1), H_{\text{Tog}}(\tau_2)].
\end{align}
We note that in the toggling frame, after removing the Rabi drive, the remaining Hamiltonian is at $O(\Omega^2/\Delta_2)$. Meanwhile, $T\sim T_g\sim 1/\Omega$ is the gate time. Therefore, $\mathcal{Z}_1\sim O(\Omega/\Delta_2)$ and $\mathcal{Z}_2\sim O(\Omega/\Delta_2)^2$. To stay consistent with the Schrieffer-Wolff expansion, we keep only $\mathcal{Z}_1$. The multiplicative error $\tilde{U}(T)\approx\exp[-i\mathcal{Z}_1(T)]$ reads
\begin{align}\label{eqn:error_unitary_Magnus_first_order}
    \tilde{U}(T) &\approx \exp\left[-i\int_0^T H_{\text{Tog}}(\tau)d\tau\right]\notag\\
    &=\exp\left[-i\left(\frac{\delta_{y}}{2}\sigma^y_{j,j+1}+\frac{\delta_{z}}{2}\sigma^z_{j,j+1}\right)\right],
\end{align}
where 
\begin{align}
    \delta_{y} &= \int_{0}^{T}\left[\lambda_j{\Omega_y}\cos\theta+\zeta(t)\sin\theta\right]d\tau,\\
    \delta_{z} &= \int_{0}^{T}\left[-\lambda_j{\Omega_y}\sin\theta+\zeta(t)\cos\theta\right]d\tau.
\end{align}
Using the BCH formula, we factorize
\begin{align}
    \tilde{U}(T) &\approx \exp\left[-i\frac{\delta_{y}}{2}\sigma^y_{j,j+1}\right]\exp\left[-i\frac{\delta_{z}}{2}\sigma^z_{j,j+1}\right],
\end{align}
where the truncation error due to discarding the commutator is at $O(\Omega/\Delta_2)^2$. We emphasize that this is consistent with the effective Hamiltonian at $O(\Omega^2/\Delta_2)$ and the Magnus expansion at $O(\Omega/\Delta_2)$.

\section*{The error parametrization}
Throughout, $X(\theta)=\exp[-i\frac{\theta}{2}\sigma^x_{j,j+1}]$ and $Y(\theta)=\exp[-i\frac{\theta}{2}\sigma^y_{j,j+1}]$ denote rotations of the $\{|j\rangle,|j+1\rangle\}$ subspace, while $Z(\delta)=\mathrm{diag}(1,e^{i\delta})$ is defined with the global phase $e^{-i\delta/2}$ factored out.

In the drive frame $\omega_d$, the total evolution reads
\begin{align}
    U(T) &= V_{\text{Tog}}(T)\tilde{U}(T),\\
    &\approx e^{-i\frac{\theta}{2}\sigma^x_{j,j+1}}e^{-i\frac{\delta_{y}}{2}\sigma^y_{j,j+1}}e^{-i\frac{\delta_{z}}{2}\sigma^z_{j,j+1}},
\end{align} 
In the special case $\theta(T_g)=\pi/2$, i.e., a $\pi/2$ rotation, the $\sigma^y$ error can be propagated through the ideal $\sigma^x$ rotation to obtain, up to a global phase,
\begin{align}
    U(T_g)=Z(\delta_y)X(\pi/2)Z(\delta_z).
\end{align}
We now consider a symmetric pulse of duration $T_g$, $\Omega_x(T_g-t)=\Omega_x(t)$, implying $\theta(T_g-t)=\frac{\pi}{2}-\theta(t)$. The $\sin\theta$ and $\cos\theta$ swap under $t\to T_g-t$, while $|\Omega|^2$ is invariant and the quadrature is antisymmetric, $\Omega_y(T_g-t)=-\Omega_y(t)$ (as holds for a DRAG quadrature $\Omega_y\propto\dot\Omega_x$). The toggling frame Hamiltonian $H_\text{Tog}(T_g-t)$ therefore maps $\sigma^y$ to $\sigma^z$ and vice versa under this time reversal. The accumulated $\delta_y$ and $\delta_z$ errors, being integrals of $H_\text{Tog}$, are thus equal, $\delta_y=\delta_z$. We arrive at the error parametrization
\begin{align}  
    U(T_g) \equiv \mathcal{X}(\pi/2) = Z(\delta)X(\pi/2)Z(\delta).
\end{align}
This parametrization holds for any symmetric pulse envelope, $\Omega_x(T_g-t)=\Omega_x(t)$, e.g., a truncated Gaussian or raised-cosine envelope, and only for $\pi/2$ rotations, $\theta(T_g)=\pi/2$, which allowed the $\sigma^y$ error to be propagated to a $\sigma^z$ phase error $\delta$. All three truncations we have made (Schrieffer-Wolff, Magnus, and BCH expansion) are at $O((\Omega/\Delta_2)^2)$, so the phase error $\delta$ is accurate to $O(\Omega/\Delta_2)$. 

In practice, the condition for $\sigma^y\to\sigma^z$ propagation $\theta(T_g)=\pi/2$ is never exactly met. Control electronics are known to output temperature-dependent pulse envelopes, which change $\Omega_x(t)$ over time; fluctuations in the qubit frequency $\delta_q$ also change the response of the qubit to the control drive, and thus $\lambda_j$; and lastly a $\sigma^x$ correction enters at $O(\Omega^3/\Delta_2^2)$ in the Schrieffer-Wolff transformation. To a good approximation, we can account for all these effects by letting 
\begin{align}
    \theta(T_g) = \frac{\pi}{2}+\epsilon,
\end{align}
where $\epsilon$ is a small correction term. The symmetric condition $\Omega_x(T_g-t)=\Omega_x(t)$ now implies $\theta(T_g-t)=\frac{\pi}{2}+\epsilon-\theta(t)$, so the time-reversal symmetry that exchanges $\sigma^y\leftrightarrow\sigma^z$ is approximately preserved,
\begin{align}
    |\delta_y-\delta_z|=\epsilon\delta + O(\epsilon^2)
\end{align}
as long as $\epsilon\delta\ll 1$. The total unitary generalizes to
\begin{align}
    \mathcal{X}(\pi/2) = Z(\delta)X(\pi/2+\epsilon)Z(\delta).
\end{align} 
Crucially, $\epsilon$ and $\delta$ are not independent. Let $\Omega_x(t)=\Omega_0(1+\eta)f(t)$ (where $\eta\ll 1$) be the actual pulse envelope; $f(t)$ is the normalized pulse envelope that satisfies the symmetric condition $f(T_g-t)=f(t)$. Both the DRAG quadrature $\Omega_y$ and the net frequency shift $\zeta$ follow
\begin{align}
    \Omega_y(t) &= -\beta\frac{\Omega_0(1+\eta)\dot{f}(t)}{\Delta_2},\\
    \zeta(t) &= \delta_q + \frac{\lambda_{j+1}^2-\lambda_{j-1}^2}{4\Delta_2}\Omega_0^2(1+\eta)^2f^2(t).
\end{align}
Thus, the phase error $\delta$ is a function of the rotation error $\epsilon$. On the other hand, the rotation error $\epsilon$ depends only on the pulse area. In other words, a calibration protocol that determines a phase-error correction (e.g., a DRAG correction) does not affect the rotation error, while a calibration protocol that determines a rotation-error correction (e.g., pulse amplitude rescaling) affects the phase error.

\section*{Closed form of DRAPE state}\label{appendix:drape_probability}
The erroneous operators are $\mathcal{X/Y}(\pi/2)=Z(\delta)X/Y[\pm(\pi/2+\epsilon)]Z(\delta)$. The backbone of the DRAPE sequence is the pseudo-identity $[\mathcal{X}(-\pi/2)\mathcal{X}(+\pi/2)]^k$. Let $c=\cos[(\pi/2+\epsilon)/2]$ and $s=\sin[(\pi/2+\epsilon)/2]$. We compute the analytical form of this pseudo-identity by first recognizing that 
\begin{align}
    M&=\mathcal{X}(-\pi/2)\mathcal{X}(+\pi/2),\\
    &= e^{2i\delta} \begin{pmatrix}
        c^2e^{-2i\delta}+s^2 & -\cos\epsilon\sin\delta \\ 
        \cos\epsilon\sin\delta & s^2+c^2e^{2i\delta}
    \end{pmatrix},\\
    &=e^{2i\delta}N
\end{align}
where $\det(M)=e^{4i\delta}$, $\det(N)=1$ and $\text{tr}(N)=2\cos^2\delta+2\sin\epsilon\sin^2\delta$. We can exploit the latter properties to calculate the characteristic polynomial $p(N)$ using the Cayley-Hamilton theorem $N^2-\text{tr}(N)N+I=0$. In fact, we can generalize the Cayley-Hamilton theorem to
\begin{align}
    N^k=\text{tr}(N)N^{k-1}-N^{k-2},\\
    N^0 = I, N^1 = N.
\end{align}
By induction, every power of $N$ is a linear combination of $N$ and $I$, $N^k=f_k N +g_kI$ for some scalar coefficients $f_k,g_k$. The scalar coefficients themselves also have their own recurrence and initial conditions
\begin{align}
    f_k = \text{tr}(N)f_{k-1}-f_{k-2},\\
    f_0=0, f_1=1,\\
    g_k=\text{tr}(N) g_{k-1}-g_{k-2},\\
    g_0=1, g_1=0.
\end{align}
This is the same recurrence structure of the Chebyshev polynomials of the second kind $S_{k}(x)$,
\begin{align}
    S_{k}(x)=2xS_{k-1}(x)-S_{k-2}(x),\\
    S_0=1, S_1=2x.
\end{align}
Using the trigonometric definition of $S_k$ and the initial conditions of $f,g$, we identify them with $f_k=S_{k-1}(\cos\phi)$ and $g_k=-S_{k-2}(\cos\phi)$, where $\cos\phi=\cos^2\delta+\sin\epsilon\sin^2\delta$. The pseudo-identity raised to the power of $k$ is
\begin{align}
    M^k = e^{2ik\delta}[S_{k-1}(\cos\phi)N-S_{k-2}(\cos\phi)I],
\end{align}
where $S_{k}(\cos\phi)=\frac{\sin[(k+1)\phi]}{\sin\phi}$.
Having reduced the pseudo-identity to a closed form, we now propagate the full DRAPE
unitary $U$ on the initial state $|j\rangle$ and read off the measured population, arriving at
Eq.~(\ref{eqn:probability_DRAPE}). First, the imperfect $|-y\rangle$ state,
\begin{align} 
    |-y\rangle = c|j\rangle -is\,e^{i\delta}|j+1\rangle
\end{align}
The final basis rotation, $\mathcal{Y}(\pi/2)=Z(\delta)Y(\pi/2+\epsilon)Z(\delta)$, gives
\begin{align}
    \langle j+1|\,\mathcal{Y}(\pi/2)=e^{i\delta} [s\langle j| + c\,e^{i\delta}\langle j+1|].
\end{align}
The population driven into $|j+1\rangle$ is $P_{|j+1\rangle}=|\langle j+1|U|j\rangle|^{2}$.
Inserting the closed form $M^{k}=e^{2ik\delta}[S_{k-1}N-S_{k-2}I]$, the accumulated
$e^{i(2k+1)\delta}$ is an unobservable global phase, so that
\begin{align}
    P_{|j\rangle}=1-|w|^{2},\\
    w=\begin{pmatrix} s & c\,e^{i\delta}\end{pmatrix}
    \left[S_{k-1}N-S_{k-2}I\right]
    \begin{pmatrix} c \\ -is\,e^{i\delta}\end{pmatrix}.
\end{align}
Eliminating $S_{k-2}$ with the Chebyshev recurrence $S_{k-2}=2\cos\phi\,S_{k-1}-S_{k}$
separates $w$ into the two independent polynomials,
\begin{align}
    w=S_{k}\,w_{I}+S_{k-1}\left(w_{N}-2\cos\phi\,w_{I}\right),
\end{align}
where $w_{I}$ and $w_{N}$ are the matrix elements of $I$ and $N$ between the prepared
and measured states,
\begin{align}
    w_{I}&=\begin{pmatrix} s & c\,e^{i\delta}\end{pmatrix}
           \begin{pmatrix} c \\ -is\,e^{i\delta}\end{pmatrix}
         = sc\left(1-i\,e^{2i\delta}\right),\\
    w_{N}&=\begin{pmatrix} s & c\,e^{i\delta}\end{pmatrix} N
           \begin{pmatrix} c \\ -is\,e^{i\delta}\end{pmatrix}.
\end{align}
Using $2sc=\cos\epsilon$, $2c^{2}=1-\sin\epsilon$, $2s^{2}=1+\sin\epsilon$, and the
identity $1-i\,e^{2i\delta}=\sqrt{2}\,(\cos\delta+\sin\delta)\,e^{i(\delta-\pi/4)}$, which follows from $e^{-i\delta}-i\,e^{i\delta}=(1-i)(\cos\delta+\sin\delta)$, a
direct evaluation gives
\begin{align}
    w_{I}&=\frac{\cos\epsilon}{\sqrt{2}}\,(\cos\delta+\sin\delta)\,e^{i(\delta-\pi/4)},\\
    w_{N}-2\cos\phi\,w_{I}&=\frac{\cos\epsilon}{\sqrt{2}}\,e^{i(\delta-\pi/4)}\notag\\
    &\times\left[-(\cos\delta+2\sin\epsilon\sin\delta)+i\sin\delta\right].
\end{align}
Both contributions carry the same phase $e^{i(\delta-\pi/4)}$, which factors out of
$w$ and cancels in $|w|^{2}$. Collecting the real and imaginary parts of the
surviving bracket,
\begin{align}
w=\frac{\cos\epsilon}{\sqrt{2}}\,e^{i(\delta-\pi/4)}\Big\{
&(\cos\delta+\sin\delta)S_k \notag\\
&-(\cos\delta+2\sin\epsilon\sin\delta)S_{k-1} \notag\\
&+\,i\,\sin\delta\,S_{k-1}
\Big\}.
\end{align}
Introducing the shorthand
\begin{align}
    u&=(\cos\delta+\sin\delta)[S_{k}-(1+\sin\epsilon)S_{k-1}],\notag\\
    v&=(1-\sin\epsilon)\sin\delta\,S_{k-1},\notag\\
    f&=\sin\epsilon\,S_{k-1},
\end{align}
the real part equals $u+v+f\cos\delta$ and the imaginary part equals
$v+f\sin\delta=\sin\delta\,S_{k-1}$, so that
\begin{align}
    P_{|j\rangle}&=1-|w|^{2},\\
    &=1-\frac{\cos^{2}\epsilon}{2}\left[(u+v+f\cos\delta)^{2}+(v+f\sin\delta)^{2}\right],
\end{align}
which is Eq.~(\ref{eqn:probability_DRAPE}); the return population follows from
unitarity, $P_{|j\rangle}=1-P_{|j+1\rangle}$. As a check, at the crossing point
$\delta=0$ one has $\cos\phi=1$ and $S_{k}(1)=k+1$, giving $u+v+f\cos\delta=1$ and
$v+f\sin\delta=0$; hence $P_{|j+1\rangle}=\tfrac{1}{2}\cos^{2}\epsilon$ independent of
$k$, the $k$-independent locus that defines $\beta_{0}$.

\section*{Fisher information of the DRAPE estimator}\label{appendix:fisher}
Each DRAPE circuit has two outcomes, $|j\rangle$ and $|j+1\rangle$, so $N$ repetitions form $N$ Bernoulli trials with success probability $P\equiv P_{|j+1\rangle}=1-P_{|j\rangle}$, with $P_{|j\rangle}$ given by Eq.~(\ref{eqn:probability_DRAPE}). The classical Fisher information for a parameter $\xi$ is therefore
\begin{align}\label{eqn:fisher_def}
    F_{\xi\xi}=N\,\frac{(\partial_\xi P)^2}{P(1-P)},
\end{align}
where $\xi\in\{\epsilon,\delta\}$ and the Cram\'er-Rao bound gives $\sigma(\hat\xi)\ge 1/\sqrt{F_{\xi\xi}}$. Since $F_{\xi\xi}$ is invariant under $P\to 1-P$, we can choose either $P_{|j\rangle}$ or $P_{|j+1\rangle}$ to proceed. Here we choose $P_{|j+1\rangle}$.

Writing Eq.~(\ref{eqn:probability_DRAPE}) as
$P=\tfrac{1}{2}\cos^2\epsilon\,(\mathcal{A}^2+\mathcal{B}^2)$ with
$\mathcal{A}=u+v+f\cos\delta$ and $\mathcal{B}=v+f\sin\delta$,
\begin{align}
    \mathcal{A}&=(\cos\delta+\sin\delta)S_k-(\cos\delta+2\sin\epsilon\sin\delta)S_{k-1},\notag\\
    \mathcal{B}&=\sin\delta\,S_{k-1},
\end{align}
its parameter derivatives are
\begin{align}
    \partial_\delta P&=\cos^2\epsilon\,(\mathcal{A}\,\partial_\delta\mathcal{A}+\mathcal{B}\,\partial_\delta\mathcal{B}),\notag\\
    \partial_\epsilon P&=-\tfrac{1}{2}\sin 2\epsilon\,(\mathcal{A}^2+\mathcal{B}^2)
    +\cos^2\epsilon\,(\mathcal{A}\,\partial_\epsilon\mathcal{A}+\mathcal{B}\,\partial_\epsilon\mathcal{B}).
\end{align}

We evaluate at the crossing $\beta=\beta_0$, i.e. $\delta=0$. There $\cos\phi=1$, so
$S_k(1)=k+1$ and $S_{k-1}(1)=k$, giving $\mathcal{A}=1$, $\mathcal{B}=0$, and
$P(\beta_0)=\tfrac{1}{2}\cos^2\epsilon$. At $\delta=0$ one has
$\partial_\delta\cos\phi=0$ (so the Chebyshev-derivative terms drop) and
$\mathcal{B}=0$, while the entire $\epsilon$-dependence of $\mathcal{A},\mathcal{B}$
enters through $\sin\delta$ or $\cos\phi$, both stationary in $\epsilon$ at
$\delta=0$. Hence
\begin{align}
    \partial_\delta P\big|_{\beta_0}&=\cos^2\epsilon\,[(k+1)-2k\sin\epsilon],\notag\\
    \partial_\epsilon P\big|_{\beta_0}&=-\tfrac{1}{2}\sin 2\epsilon,
\end{align}
the latter independent of $k$: at the crossing the population is
$\tfrac{1}{2}\cos^2\epsilon$ for every $k$. Substituting into
Eq.~(\ref{eqn:fisher_def}),
\begin{align}
    F_{\delta\delta}(\beta_0)&=N\,\frac{2\cos^2\epsilon\,[(k+1)-2k\sin\epsilon]^2}
    {1-\tfrac{1}{2}\cos^2\epsilon}\\
    F_{\epsilon\epsilon}(\beta_0)&=N\,\frac{2\sin^2\epsilon}{1-\tfrac{1}{2}\cos^2\epsilon}.
\end{align}
In the limit of $\epsilon$ goes to zero,
\begin{align}
    F_{\delta\delta}(\beta_0) \xrightarrow{\ \epsilon\to0\ }\ 4N(k+1)^2,\label{eqn:Fdd}\\
    F_{\epsilon\epsilon}(\beta_0) \xrightarrow{\ \epsilon\to0\ }\ 4N\epsilon^2.\label{eqn:Fee}
\end{align}
Because the DRAG sweep enters only through $\delta=\chi(\beta_0-\beta)$
[Eq.~(\ref{eqn:the_DRAPE_slope})], we have $\partial_{\beta_0}P=\chi\,\partial_\delta P$
and therefore
\begin{align}
    F_{\beta_0\beta_0}=\chi^2 F_{\delta\delta}\ \xrightarrow{\ \epsilon\to0\ }\ 4N\chi^2(k+1)^2 .
\end{align}
The corresponding Cram\'er-Rao bounds at the crossing are, to leading order in
$\epsilon$,
\begin{align}
    \sigma(\hat\beta_0)&\ge\frac{1}{2|\chi|\sqrt{N}\,(k+1)},\\
    \sigma(\hat\epsilon)&\ge\frac{1}{2\sqrt{N}\,|\epsilon|}.
\end{align}
The phase-correction variance scales as $(k+1)^{-2}$; since $k$ and $N$ are
independent, $\hat\beta_0$ attains Heisenberg scaling in the amplification number
$k$. The rotation-error information $F_{\epsilon\epsilon}(\beta_0)$ carries no $k$
dependence, so $\hat\epsilon$ saturates the standard quantum limit.

The behavior away from the crossing follows from the same expansion. To leading
order in $\epsilon$, $\partial_\epsilon P\simeq-\epsilon-2k\delta$, which vanishes at
$\delta=-\epsilon/(2k)$: $F_{\epsilon\epsilon}$ has a null displaced only
infinitesimally from $\beta_0$, the point at which the sign of $\epsilon$ is
unresolvable. On either side $F_{\epsilon\epsilon}$ rises to a peak set by the first
extremum of $S_k(\cos\phi)$, located at $|\delta|\propto 1/(k+1)$; the peak thus
migrates toward $\beta_0$ as $k$ increases while its height remains fixed, whereas
$F_{\delta\delta}$ is maximal at $\beta_0$. These features are confirmed numerically
in Fig.~\ref{fig:drape}(b).

\end{document}